# Real-time steerable frequency-stepped Doppler Backscattering (DBS) System for local helicon wave electric field measurements on the DIII-D tokamak

S. Chowdhury[1, a)], N. A. Crocker[1], W. A. Peebles[1], R. Lantsov[1], T. L. Rhodes[1], L. Zeng[1], B. Van Compernolle[2], S. Tang[2], R. I. Pinsker[2], A. C. Torrezan[2], J. Squire[2], R. Rupani[2, b)], R. O'Neill[2] and M, Cengher[3]

[1]*Physics and Astronomy Department, University of California Los Angeles, California 90095, USA*
[2]*General Atomics, P.O. Box 85608, San Diego, California 92121, USA*
[3]*MIT Plasma Science and Fusion Center, Cambridge, Massachusetts 02139, USA*

a)  Author to whom correspondence should be addressed: schowdhury@physics.ucla.edu.
b)  The author is now affiliated with VIE Technologies, San Diego, California.

A new frequency-stepped Doppler backscattering (DBS) system has been integrated into a real-time steerable electron cyclotron heating launcher system to simultaneously probe local background turbulence (f < 10 MHz) and high-frequency (20 – 550 MHz) density fluctuations in the DIII-D tokamak. The launcher allows for 2D steering (horizontally and vertically) over wide angular ranges to optimize probe location and wavenumber response. The vertical steering can be optimized during a discharge in real time. The new DBS system employs a programmable frequency synthesizer with adjustable dwell time as a source to launch either O or X-mode polarized millimeter waves. This system can step in real-time over the entire E-band frequency range (60 – 90 GHz). This combination of capabilities allows for diagnosis of the complex internal spatial structure of high power (> 200 kW) helicon waves (476 MHz) injected from an external antenna during helicon current drive experiments in DIII-D. Broadband density fluctuations around the helicon frequency are observed during real-time scans of measurement location and wavenumber during these experiments. Analysis indicates that these broadband high-frequency fluctuations are a result of backscattering of the DBS millimeter-wave probe beam from plasma turbulence modulated by the helicon wave. It is observed that background turbulence is effectively locally 'tagged' with the helicon wave electric field, forming images of the turbulent spectrum in the overall density fluctuation spectrum that appear as high frequency sidebands of the turbulence. These observations of background turbulence and high-frequency fluctuations open up the possibility of monitoring local helicon wave amplitude by comparison of the high-frequency signal amplitude to the simultaneously measured background turbulence. In combination with the real-time measurement location and wavenumber scanning capabilities (offered by real-time frequency-stepping and steering), this allows rapid determination of the spatial distribution of the helicon wave power during steady-state plasma operation. In the long term, such measurements may be used to validate predictive modeling (GENRAY [1] or AORSA [2]) of helicon current drive in DIII-D plasmas.

## 1. Introduction

The use of millimeter wave Doppler backscattering (aka Doppler reflectometry) to diagnose low-frequency plasma turbulence has become standard in fusion research plasmas because of the high spatial (sub-centimeter) and temporal resolution (sub-millisecond) measurement capabilities [3,4]. Doppler backscattering (DBS) typically uses a millimeter wave beam to perform wave-number resolved, local measurements of plasma density fluctuations ($\tilde{n}$) that satisfy the Bragg scattering wavenumber relation, $\boldsymbol{k}_{\tilde{x}} = -2\boldsymbol{k}_i$, where the subscripts $\tilde{x} = \tilde{n}$ or $\tilde{b}_\parallel$ refer to the probed fluctuation and $i$ to the incident millimeter waves at the measurement location. Millimeter waves backscattered from the DBS beam by turbulent density fluctuations near millimeter wave cutoff surfaces carry local information about the turbulence [5-7]. Turbulence at different scale lengths has already been studied in DIII-D (e.g. [8,9], NSTX [10,11], MAST-U [12-14]), and other tokamaks using this technique. The $\boldsymbol{E} \times \boldsymbol{B}$ velocity and wave electric field ($\boldsymbol{E}$) amplitudes are inferred from the measured Doppler shift in the quadrature spectrum and the scattered power provides a relative measurement of the density fluctuation level. Millimeter-wave diagnostics such as DBS have more recently also been applied to probe radio or high frequency waves in fusion plasmas. Various high-frequency waves, including ion cyclotron emission [15], high frequency Alfvén eigenmodes [16], and lower hybrid waves [17] have been investigated.

The new DBS system described here is based on a previous prototype E-band DBS system which demonstrated the ability to observe a broad range of radio frequency plasma waves [16]. The new system has also demonstrated a capability to observe fluctuations near the frequency of externally injected high power helicon waves (476 MHz). Previously the observation involved direct scattering of the millimeter waves from broadband density fluctuations around the helicon frequency. These broadband helicon fluctuations were thought to be generated via interaction of the injected helicon wave with turbulence. One possible interaction is modulation of the helicon antenna coupling by turbulence in the plasma scrape off layer (SOL) in front of the antenna [18]. This paper will discuss observations of high frequency fluctuations with the new system that result from a distinctly different mechanism: direct backscattering from turbulent density fluctuations modulated by the injected helicon wave. As explained below, this mechanism will allow the spatial distribution of the helicon wave to be routinely generated with minimal assumptions.

This paper also describes integration of the DBS diagnostic with an electron cyclotron heating (ECH) overmoded waveguide propagation system [19, 20] which allows the launched beam to be steered both toroidally and poloidally. The integration utilizes a high-power overmoded waveguide switch which allows either ECH heating power to be launched or DBS experiments to be performed on the fusion plasma. The waveguide switch selects either ECH or mm-wave to propagate to the plasma using a pneumatically controlled linear actuator to control a mirror. Insertion of the mirror allows the DBS system to couple to the plasma, whereas removal of the mirror allows free propagation of the megawatt level ECH to the plasma. The development and demonstration of this technique for interfacing the DBS system with plasma establishes a model for potentially adding new diagnostic capabilities to future burning plasma facilities which employ ECH. Among the details to be discussed is the waveguide switch that allows either the ECH source or DBS diagnostic to couple millimeter waves into the overmoded waveguide transport system, and the quasi-optical technique utilized to couple the DBS source radiation into the waveguide switch.

This paper reports these new design improvements and describes results from recent helicon experiments in the DIII-D tokamak. Both frequency stepping and dynamic steering during single plasma discharges are used to efficiently extract a wealth of fluctuation data with the DBS system. These results establish the DBS system as a sensitive, flexible diagnostic, with fine temporal and spatial resolution as well as broad spatial and wave number coverage, to characterize a broad spectral range ($f$ = 20 – 550 MHz) of plasma waves present in a single plasma discharge.

The primary aim of the upgraded DBS system is to measure helicon wave-field amplitude and spatial structure and any interaction with simultaneously measured turbulence. This supports objectives of the DIII-D research program to establish helicon waves as an efficient off-axis current drive system to achieve the goal of steady-state reactor operation. [21 – 23]. Measurements of helicon wave-field amplitude and spatial structure with the new DBS will be used to validate theory and code predicted (GENRAY [1] and AORSA [2, 18]) performance of helicon wave heating and current-drive in terms of propagation and absorption inside the plasma.

Details of the improved system design are discussed in Sec. 2. Section 3 describes scattering measurements in recent DIII-D plasmas, including results demonstrating frequency stepping and beam steering within individual discharges. Section 4 discusses the physics importance and summarizes the significance of the helicon measurements achieved in order to support the helicon current drive program at DIII-D.

**2. SYSTEM DESIGN AND MAJOR UPGRADES**

The system reported here is upgraded in several ways relative to the prototype system [16], resulting in improved sensitivity and spatial and wavenumber coverage, especially for helicon wave measurements. Figure 1 shows the overall structure of the new DBS. The new system retains the quadrature configuration of the prototype, with the ability to change the operating frequency of the millimeter wave probe beam. However, the millimeter wave source has been modified to allow for real-time tuning over the entire 60–90 GHz E-band range. In addition, several modifications to the receiver circuit improve the system sensitivity, including hardening against pickup of stray 476 MHz helicon radiation. Significantly, the new DBS also differs from the prototype in how it interfaces with plasma. The new DBS system is quasi-optically coupled to a waveguide switch which integrates the DBS into a corrugated waveguide. This waveguide normally transports ECH power to a launcher, or steering assembly, for directing the power into the plasma [19, 20]. The DBS uses the toroidal and poloidal aiming capabilities of the ECH launcher as well as the ECH optics focusing capability. The ECH launcher has a much wider

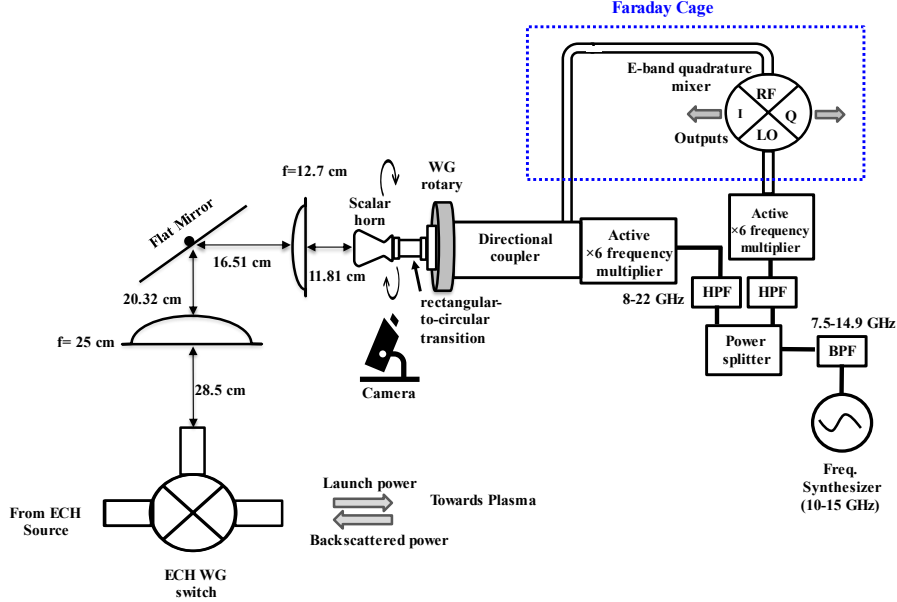

FIG. 1. Diagram showing mm-wave circuit with quasi-optics setup for DBS measurements. A tunable frequency synthesizer (10–15 GHz) is frequency-multiplied into the E-band (60 – 90 GHz) range to probe plasma density fluctuation at different cut-off locations. A Faraday shield with band-pass and low pass filtered penetrations encloses the receiver circuit to minimize signal contamination from pickup of stray RF radiation. A waveguide rotary joint before the scalar horn is used to select launch mm-wave polarization remotely. A combination of two lenses (one to collimate and the other to focus) and a mirror are used to couple the DBS system to a waveguide switch in an overmoded corrugated waveguide leading to the plasma. The waveguide switch selects whether the ECH source or DBS system connects to this corrugated waveguide.

range of steering angles than the steering assembly used by the prototype DBS [16].

An overview of system in Fig. 1 shows many of the modifications to the millimeter wave circuit, including changes to the millimeter wave source and the enclosure of receiver circuit with Faraday shielding. The range of millimeter wave operating frequencies is expanded to the full E-band (60 – 90 GHz) by replacing a pair of phase-locked dielectric resonator oscillator (PLDRO) sources featured in the prototype with a remotely programmable frequency synthesizer (Mercury DS-3002). The synthesizer can produce frequencies over a range of 0.1–20 GHz with low phase noise. The synthesizer is used to produce frequencies in the 10 – 15 GHz range that are multiplied to the 60 – 90 GHz range by active ×6 active frequency multipliers (Eravant, SFA-603903620-12SF-E1). Laboratory tests show high-frequency stability and improved signal-to-noise compared to the prototype system [16] when using the synthesizer.

In contrast, with the prototype system, the new DBS system uses active ×6 frequency multipliers in place of ×4 multipliers. The change in multiplication factor in the multipliers was made because the programming circuitry of the synthesizer is anticipated to be vulnerable to neutron radiation from DIII-D plasmas. To mitigate this danger, the synthesizer is also placed approximately 5 meters from the tokamak, outside a concrete shield wall that largely contains the D-D fusion neutrons. This minimizes the possibility of neutron damage to the programmable circuitry, as well as random bit-flips of the program memory and logical control circuitry of the synthesizer. The synthesizer output is transmitted to the multipliers via a low-loss (~0.15dB/ft) 30-feet coax RF cable (model: Teledyne True Blue 420) through a penetration in the shield walls. The cable is well shielded against pickup of any stray helicon radiation from the high-power antenna that leaks into the area around the tokamak.

The programmability of the synthesizer also allows for efficient exploitation of DIII-D discharges, which can last up to 10 seconds. The operating frequency can be stepped during a discharge, allowing probing of multiple radial locations during a single discharge (as compared to multiple repeat discharges with a fixed source like in the prototype system [16]). In sweep mode, the synthesizer can step in increments as small as 1 Hz through a chosen range with a 1 millisecond settling time after each step. The dwell time, which applies to all steps, can be set within a range from 0.5 ms to many hundreds of milliseconds. In 'list' mode, it can step through a list of up to 5000 different frequencies, with a settling time of 200 µs after each step, and a different dwell time for each step. The flexibility of the programming allows for both fine and rough frequency scans. The synthesizer accepts an external trigger that controls timing of the synthesizer program execution. The trigger is taken from the timing system that controls the DIII-D discharge control and data acquisition systems in order to synchronize DBS operation with these systems. Finally, the synthesizer

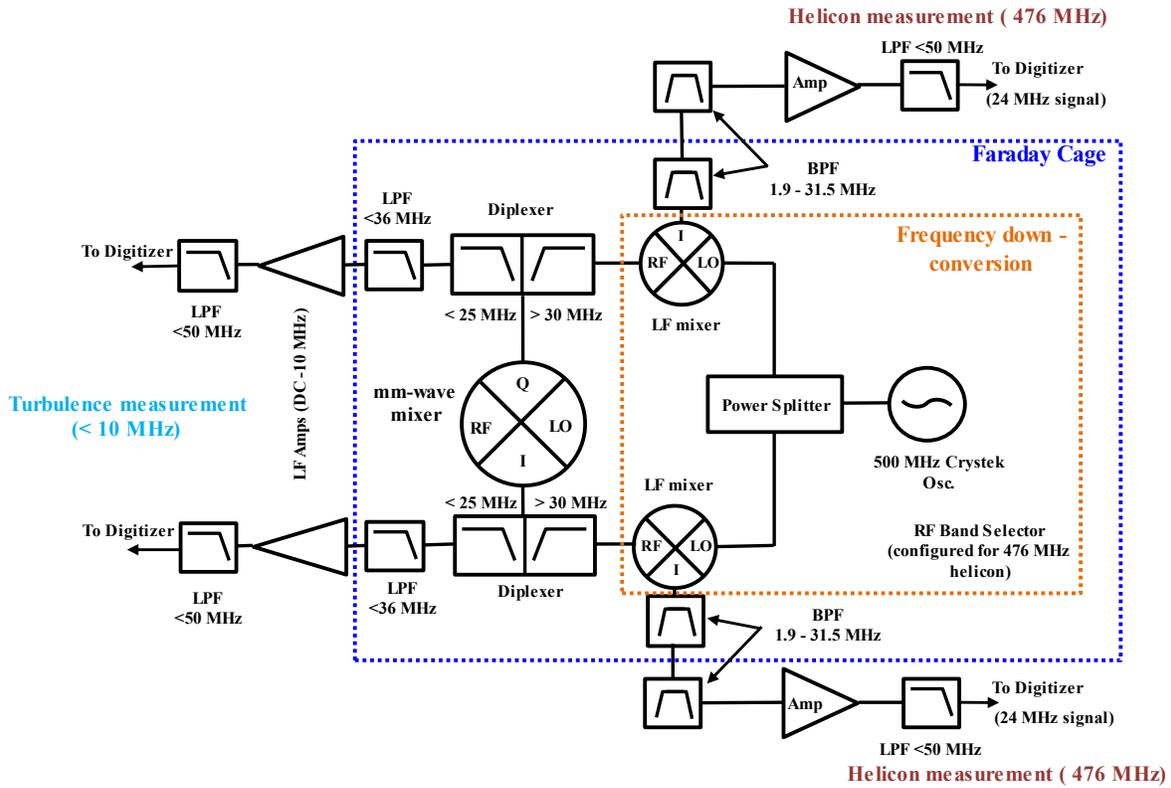

FIG. 2. Receiving circuit with frequency down conversion setup for high frequency fluctuation measurements in helicon frequency range (f = 476 MHz) and low-frequency turbulence in separate channels. A 500 MHz crystal oscillator down-converts the high frequency signal around 476 MHz to around 24 MHz. The signals are then digitized by a high frequency digitizer. (The lower frequency turbulence signal (<10 MHz) does not require the high frequency capability of the digitizer, but the use of a single multi-channel digitizer simplifies the system.)

is remotely programmed via a USB relay and a USB interface.

As with the prototype DBS system, the receiver circuit separates low-frequency and high-frequency fluctuation signals using diplexers at the millimeter wave mixer in-phase (I) and quadrature (Q) signal output ports. New amplifiers (Analog Module 351A-1-50-NI) are used for lower noise with more uniform gain and higher saturation voltage throughout the required frequency range (f ≲ 10 MHz). Prior to amplification, the high-frequency signals are down-shifted to lower frequency by mixing with a 500 MHz crystal oscillator in a Mini-circuits low-frequency mixer. As with the previous system, all signals from the DBS system are recorded with a fast-sampling digitizer (100 MS/s) with high input bandwidth, $f < 100$ MHz, limited to $f < 50$ MHz by external anti-alias filters.

A second, more significant, upgrade is to improve rejection of unwanted external pickup of 476 MHz radiation from the high-power helicon antenna. A 1.5-mm thick aluminum Faraday shielding box (Fig 2) now encloses the millimeter-wave quadrature mixer, diplexers, crystal oscillator and RF mixers. Coax cabling along with RF filters to reject 476 MHz are used for all signals and power supply cables penetrating the shield box. For the low-frequency signals, a 36 MHz low pass filter is used (Mini-Circuits SLP-36+), while for the down-shifted high-frequency signals, a 1.9 – 31.5 MHz high pass filter is used (Mini-Circuits ZABP-16+). A second RF pickup rejection technique is the insertion of 8–24 GHz bandpass filters (Mini-Circuits ZHSS-8G-S+) at the source side of active multiplier inputs in the millimeter-wave circuit (Fig. 1). The active multipliers have coaxial inputs that could otherwise allow 476 MHz pickup along the input line to enter the multiplier and potentially modulate the output LO and probe beam millimeter waves. Laboratory tests show that these combined arrangements improve RF pickup rejection by ~34 dB.

The millimeter wave circuit couples (Figs. 1, 3 and 4) into an electron cyclotron heating system in order to take advantage of the high-power millimeter wave beam steering assembly, or launcher, for steering the DBS probe beam. The launcher system [19, 20] contains a paraboloidal fixed mirror which focusses radiation from a 60.325 mm corrugated overmoded waveguide onto a steerable flat mirror. The focusing mirror is designed so that radiation propagating in the $HE_{11}$ mode in the waveguide, after emerging from the waveguide aperture and reflection from the focusing mirror, forms a Gaussian beam in the far field. At 110 GHz the focusing mirror produces a 13 cm diameter spot size beam at a distance of 1 m from the radiating waveguide aperture [24], which is well inside the core plasma.

The steerable mirror is driven by a two-axis motorized system for poloidal and toroidal beam steering. The poloidal steering angle can range from 29º above horizontal to 43º below horizontal. The toroidal steering angle can range from 34º right of the inward major radial direction to 28º left, when viewed from above. The mirror angles can be set before each plasma discharge and held fixed throughout, in a mode of operation similar to steering for results reported in [16]. However, a control system for the motor that sets the poloidal steering angle can also be programmed to change the angle throughout the discharge, as quickly as 40 -50°/sec, allowing significant angle changes during the course of a discharge. With this steering capability, the DBS system can cover a fluctuation wavenumber range of nearly $|k_\theta| = 0 - 25$ cm$^{-1}$. This in-shot steering capability is demonstrated in experiments reported here. The toroidal steering angle is also set by a motor, but the control system for the motor can only set the angle before each discharge and hold it steady throughout. To prepare for an experiment, ray tracing is performed with the GENRAY [1] or TORAY [25] millimeter wave ray tracing codes to determine the optimum steering angles for the target wavenumber, using a model density profile and equilibrium magnetic field of the anticipated discharge (typically based on a similar discharge which is being recreated).

As mentioned earlier, in order to interface with the ECH waveguide propagation system, the millimeter wave circuit is quasi-optically coupled via a T-shaped waveguide switch [26] into a circular corrugated waveguide (Figs. 1, 3, and 4) leading to the plasma. This corrugated waveguide normally feeds power directly from a gyrotron to the ECH launcher. It has a 31.75 mm inner diameter at the location of the switch, tapering up to the 60.325 mm inner diameter waveguide of the launcher assembly approximately 1 meter from the launcher. All legs of the switch are also corrugated, with a diameter of 31.75 mm. Pneumatic actuators are utilized to insert a gold-coated metal mirror into the junction of the switch allowing a 90° reflection of the DBS probe radiation into the corrugated waveguide leading towards the plasma. Withdrawal of the mirror leaves the corrugated waveguide open for high-power gyrotron ECH radiation to pass into the plasma. Interlocks control the operation of the switch and an operation safety protocol is implemented to protect both the ECH and DBS systems. It is important to mention that with this arrangement DBS diagnostic can operate only when the waveguide is not used by the ECH gyrotron. The waveguide switch [27] design ensures negligible leakage from the ECH input leg to the DBS input leg when in position for ECH operation.

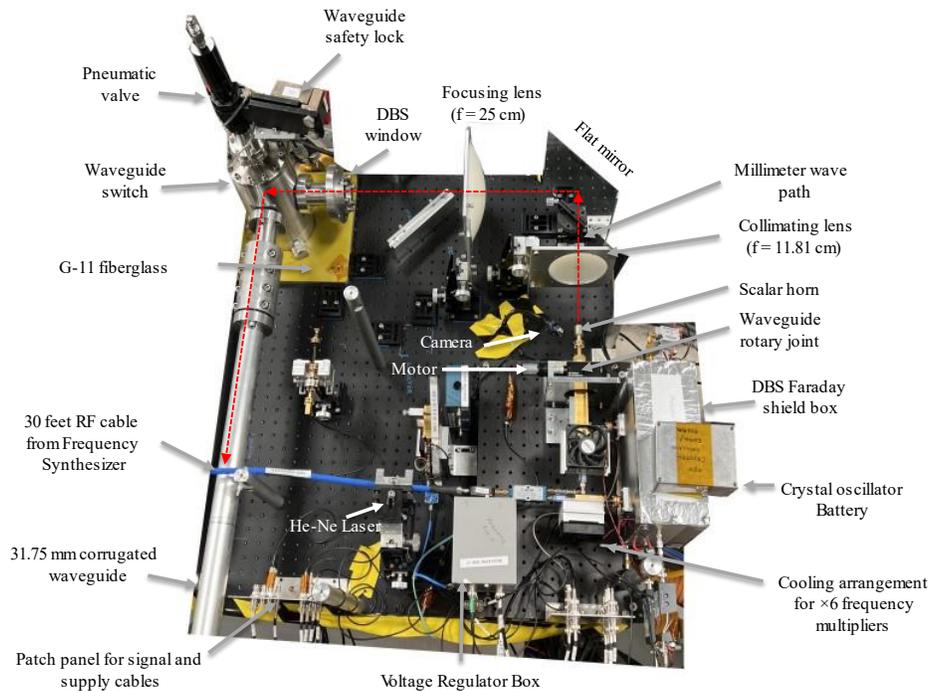

FIG. 3. Photo of optical breadboard showing quasi-optics, waveguide rotary joint, DBS beam path, and other associated components of the millimeter wave circuit. Photo also shows the 31.75 corrugated waveguide leading to the plasma and the waveguide switch coupling the DBS beam into waveguide.

The quasi-optical assembly for coupling the millimeter wave circuit to the switch consists of a scalar horn, a pair of lenses and a fully adjustable mirror (Figs. 3 and 4). DBS launch radiation from the horn is focused by a hyperbolic high-density polyethylene (HDPE) lens towards a second hyperbolic lens, which focuses the radiation to the entrance aperture of the waveguide switch to facilitate high-quality coupling to an HE$_{11}$ mode. Testing indicates that this assembly effectively couples millimeter waves into the HE$_{11}$ mode throughout the full E-band range.

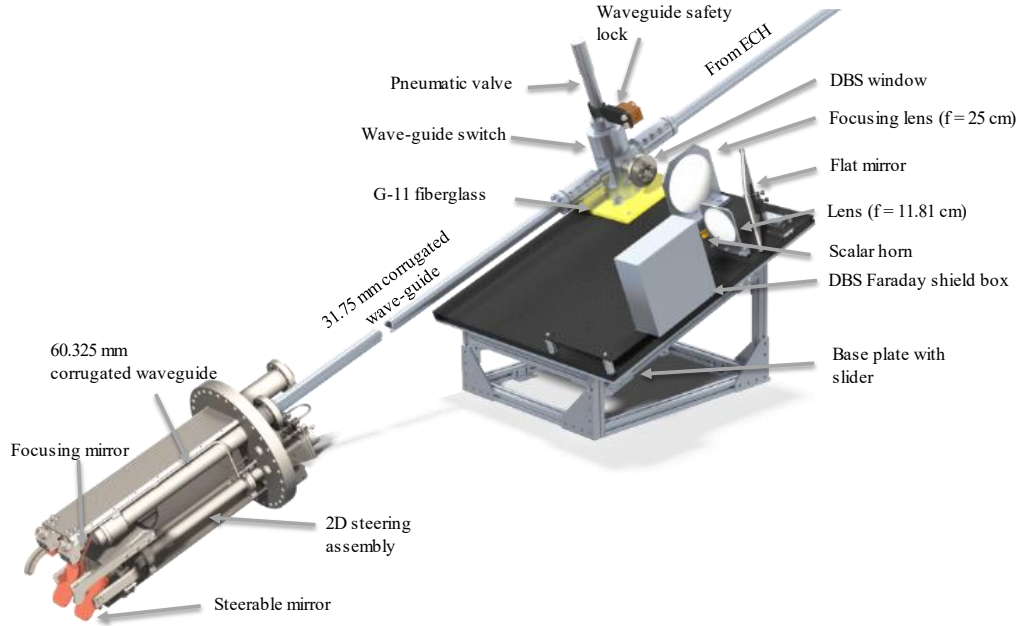

FIG. 4. 3D schematic illustrating the overall configuration of the DBS system. The breadboard is positioned ~3 m away from DIII-D vacuum vessel on a translatable frame isolated from the vessel ground. The translating frame allows for changes in position due to thermal expansion of the DIII-D vacuum vessel and ECH waveguides.

For compactness, the quasi-optical assembly is laid out on a breadboard with a mirror between the two lenses to redirect the path of the millimeter wave beam. The lenses and mirror are mounted on stages with various degrees of translatability to permit alignment of the components. The quasi-optics and switch are aligned on the DBS breadboard in the laboratory before the breadboard is positioned to integrate the switch into the ECH waveguide at the DIII-D facility. The procedure for alignment involves first mounting the switch on the breadboard, and then using a laser to align the quasi-optical system to ensure that it will focus a beam on the center of the switch input leg aperture. The switch is then removed, and the quasi-optical system is tuned to focus the beam to a suitable beam waist at the location of the aperture. The beam profile is measured vertically and horizontally. After tuning, the beam dimension at the aperture is small enough across the 60 – 90 GHz operating range of the DBS system to efficiently couple to the $HE_{11}$ mode [28]. At 63 GHz the beam diameter is 1.9 cm, while at 88.8 GHz it is 1.6 cm. The beam diameter is measured between points of the intensity profile that have $1/e^2$ of the peak intensity. After the switch is mounted, beam profiles for different operating frequencies are measured vertically and horizontally at different distances from the output leg aperture in the laboratory. Gaussian profiles are verified for a broad range of frequencies (63 – 87 GHz).

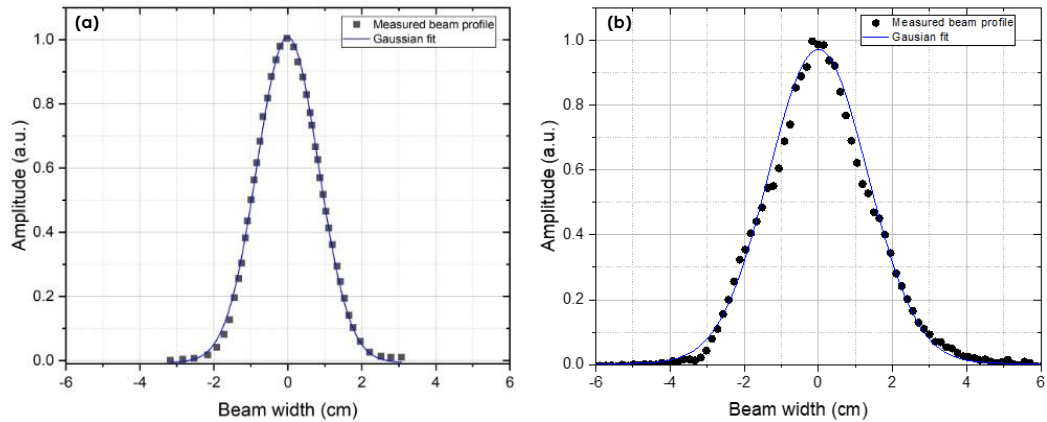

FIG. 5. Beam profiles perpendicular to the propagation direction at 78GHz for two different cases, before and after the aligned DBS system is integrated in the ECH waveguide at the DIII-D facility: (a) 10" from the waveguide switch output in the laboratory and (b) 10.5" from the ECH launcher steering mirror inside the DIII-D vacuum vessel after propagation through ECH waveguide an focusing and steering mirrors.

After the switch is integrated into the ECH waveguide at the DIII-D facility, horizontal and vertical profiles are again obtained inside the DIII-D vacuum vessel. For the in-vessel measurements the 78 GHz signal couples into the waveguide switch, propagates through ~ 3m of corrugated waveguide to the launcher, and is launched into the vacuum vessel by the ECH beam focusing and steering system. Figure 5a shows a beam profile at 78 GHz measured in the laboratory at 10" away from the waveguide switch output with the aligned quasi-optical system and switch mounted on the breadboard. Profile measurements of the beam emerging from the launcher within the DIII-D vessel are shown in Fig. 5b. The beam is poloidally steered to be horizontal and toroidally steered in the negative major radial direction. The profile is measured at a distance of 10.5" from the steering mirror. Deviation from a Gaussian structure in Fig 5b is due to reflections from the detector setup, table, and tokamak walls. These profiles indicate reasonably good coupling from the quasi-optics to the corrugated waveguide and steering system into the plasma.

The quasi-optical coupling scheme for the new DBS delivers several benefits. One benefit is that the millimeter waves propagate in the corrugated waveguide as an $HE_{11}$ mode [29] with a low loss between the waveguide switch input and the ECH launcher. Another benefit is that the radiation propagating in the $HE_{11}$ mode benefits from the launcher focusing mirror design in that it creates a reasonably good beam profile in the far field (Fig. 5b). A third benefit is spatial filtering of radiation transmitted from the plasma to the millimeter wave circuit, which enforces spatial localization of the measurement and limits noise from plasma electron emission. Different waveguide modes produce different antenna patterns in plasma with different spatial footprints. Scattered millimeter waves from the DBS probe beam, as well as electron cyclotron emission from the plasma, that match the antenna patterns of other modes besides the $HE_{11}$ can couple to these modes in the corrugated waveguide and propagate back towards the millimeter wave circuit. The quasi-optical assembly preferentially couples the $HE_{11}$ over other modes into the millimeter wave circuit, effectively restricting the antenna pattern for the DBS system to that of the $HE_{11}$ mode and reinforcing the spatial localization of the DBS measurement. The restriction of the antenna pattern also reduces noise in the measurement by minimizing the collection of excessive electron cyclotron emission.

The choice to couple the DBS system to the ECH system via the waveguide switch poses mechanical and electrical challenges. The first challenge is to electrically isolate the DBS system from the ECH waveguide. This is accomplished using quasi-optical coupling. However, this creates a mechanical challenge in the process, by mechanically decoupling the DBS system from the switch. The waveguide undergoes thermal expansion and contraction during experimental operations which could translate the switch by up to ~ 0.5cm, potentially misaligning the quasi-optical assembly with the switch. The solution adopted here is to attach the switch to a 0.5" thick plate of G-11 fiberglass laminate that also connects to the optical breadboard on which the DBS system and quasi-optical assembly are mounted (Figs. 3 and 4). The insulating character of the G-11 plate ensures that electrical isolation is maintained. The breadboard is mounted via rails on a support frame that is free to translate together with the switch during thermal expansion and contraction of the waveguide and vacuum vessel (Fig. 4). The rails are necessarily angled away from horizontal for correct alignment with the waveguide. To avoid burdening the switch with the weight of the breadboard and DBS system, the breadboard and switch are attached to the support frame via a cable and pulley system to a spring underneath the breadboard (not visible in Fig. 4) that attaches to the support frame.

A new method of polarization control has also been implemented for the new DBS system using a waveguide rotary joint (Spinner BN 636282) which coaxially joins two rectangular waveguide sections and allows them to independently rotate freely around a shared axis (Fig 1 and 3). Millimeter waves in fundamental mode are transmitted between the waveguide sections, effectively spatially reorienting the wave electric field direction when the waveguide sections are rotated to have different orientations. Typical insertion loss within this rotary joint is ~1dB (0.2 dB variation over 360° of rotation). One port of the rotary joint attaches to the millimeter wave circuit while the other port attaches to the scalar feed horn antenna (Quinstar QSH-E2500) via a rectangular to circular W/G transition. The rotary joint can be arbitrarily rotated to select the angle of the rectangular port of the transition feeding the scaler horn, allowing the polarization of the radiation emerging from the horn antenna to be adjusted appropriately relative to the edge pitch of the plasma magnetic field to couple to either X-mode or O-mode. The orientation of the rotary joint is set using a motor driven by a remotely controllable power supply. The polarization can be rotated by 90° in less than a minute, making changes between X and O-mode between discharges straightforward. A camera is used to monitor the position of the rotary joint while the motor drives the rotation.

# 3. SYSTEM PERFORMANCE, FREQUENCY STEPPING AND STEERING RESULTS

The frequency stepping and steering capabilities of the DBS are demonstrated in several different DIII-D discharges. This has allowed the turbulence, Alfvén wave type instabilities and externally launched helicon waves to be studied as a function of location and probed wave number. The radial location of DBS measurements is limited by the cutoffs. In the case of an X-mode DBS beam polarization, the cutoffs depend on both the plasma density as well as B-field value. For example, with B = 2T, in X-mode the range of density is $n_{e0} \sim 4\times10^{12}$ cm$^{-3}$ to $> 10^{14}$ cm$^{-3}$ for f$_{DBS}$ = 60 to 90 GHz. Figure 6 shows turbulence measurements obtained while stepping the operating frequency in a constant density, L-mode discharge in DIII-D with plasma current and toroidal magnetic field of I$_P$ = 1 MA and B$_T$ = 2T, respectively. Neutral beam heating with ~3MW is utilized and a peak density, $n_{e0} \sim 2\times10^{13}$ cm$^{-3}$ is maintained. Measurements are obtained using the synthesizer programmed to cycle sequentially through DBS launch frequencies of 60, 66, 72 and 78 GHz (with 400 ms dwell time at each frequency). The polarization of the millimeter probe beam is X-mode, and the steering mirror is fixed to a toroidal angle of 0° and a poloidal angle relative to normal incidence of ~1.5°, which is suitable for probing lower k fluctuations. This allows the DBS beam to probe from the edge of the plasma (ρ ~ 1) to the inner core (ρ = 0.45) and measure the Doppler shift of the turbulence, which is predominantly caused by plasma E×B rotation. Here, ρ is a flux surface coordinate giving the square root of normalized toroidal flux enclosed by the flux surface and ρ = 1 corresponds to the last closed flux surface. The quadrature spectrum of turbulent density fluctuations measured by the low-frequency channel as the system steps sequentially between operating frequencies is shown in Fig. 6a while line average plasma density remains approximately constant (Fig. 6b). The quadrature spectrum is the spectrum of the complex valued representation of the electrical field of the scattered radiation, $E = I + iQ$, where $I$ and $Q$ are the low-frequency in-phase and quadrature signals. As the synthesizer steps from low to higher operating frequencies, the DBS probed location moves from edge to core, and a peak in the spectrum is observed to shift to higher frequencies as ρ decreases. The peak frequency is the Doppler shift frequency (Fig. 6a) $f_D$, where $2\pi f_D = \mathbf{k} \cdot \mathbf{v}$, $\mathbf{k}$ is the probed wave number and $\mathbf{v}$ is the turbulence lab frame phase velocity. Turbulence lab frame phase velocity is typically dominated by the E×B velocity, $\mathbf{v}_{E\times B} = \mathbf{E} \times \mathbf{B}/B^2$ for DIII-D plasmas with strong torque from beam injection such as considered here. Raytracing with the GENRAY code [1] shows (Fig. 6c) the ray trajectory for each frequency, from which the measurement location (the location of deepest penetration) is determined. The ray tracing shows an increase in the probed wavenumber as radial location ρ decreases. (Fig. 6d). The plasma radial electric field (E$_r$) is estimated from the probe wavenumber (as predicted by GENRAY, Fig. 6d) as well as DBS measured Doppler shift in frequency (f$_D$ from Fig. 6a). The E$_r$ profile shows a typical [30] behavior characterized by elevated values at mid-radius, gradually decreasing to zero at the plasma edge. This is consistent with typical plasma conditions and E$_r$ measured by other diagnostics [30, 31] at DIII-D. This is attributed to the interplay between ion orbit loss and the radial force balance expressed through Ohm's law [30]. The strong ion orbit loss in the outer mid-radius region creates an inward-pointing radial electric field to confine the remaining plasma, leading to the observed peak at ρ ~ 0.6. The radial electric field then drops to zero at the very edge (ρ > 0.9) because of returning particles from the scrape-off layer. Note that the difference in magnitude of $E_r$ shown here and in ref [30] can be attributed to the poorer L-mode confinement in the present study, as opposed to the better H-mode confinement discussed in the referenced work.

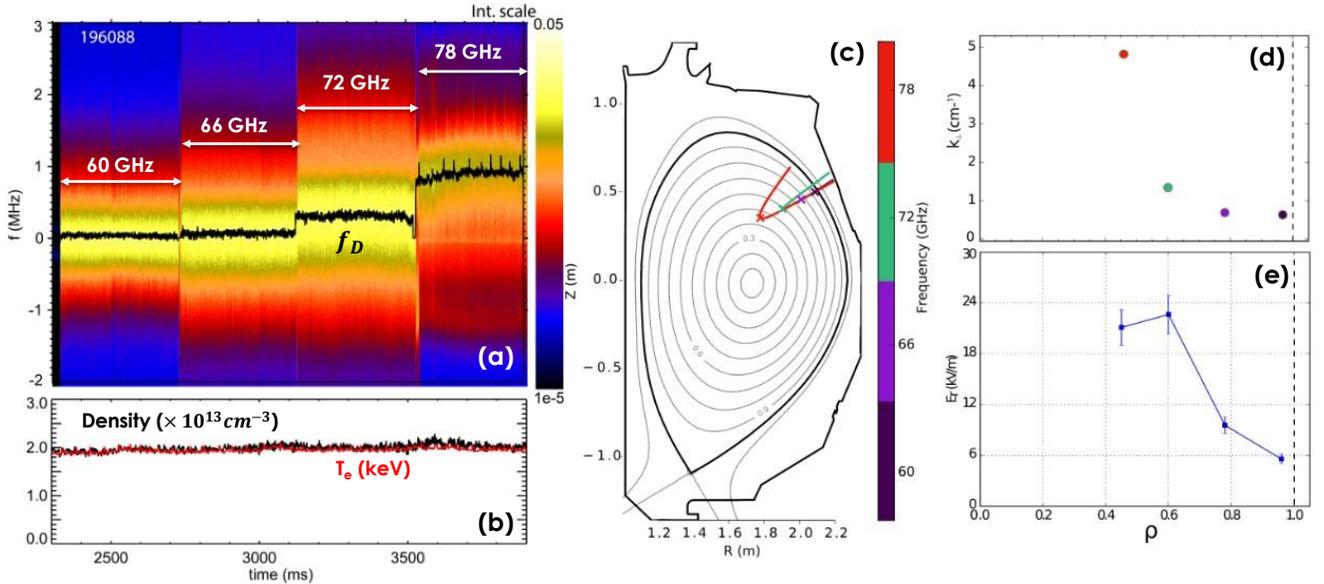

FIG. 6. (a) Quadrature frequency spectrum vs time (log$_{10}$ ñ scale) for low-frequency turbulence channel illustrating frequency steeping (60, 66, 72, 78 GHz) and mean DBS Doppler shift (f$_D$) solid black line, (b) during an approximately constant density and temperature L-mode plasma. (c) GENRAY ray tracing for these frequencies overlaid on flux contours labeled according to ρ showing radial coverage of measurement location (ρ ~ 0.45 – 1), which is point of deepest ray penetration. (d) Wavenumber (k$_\perp$) range of backscattered radiation at measurement location, and (e) estimated radial electric field (E$_r$) at measurement location using mean DBS Doppler shift (f$_D$) from (a) and probed wavenumber (k$_\perp$) from (d).

A second discharge (Fig 7) illustrates the measurement of mid frequency fluctuations (f ~ 5 – 5.5 MHz) that resemble global Alfvén eigenmodes (GAE), as described in [32] and [33]. The discharge for this case is a high-power (~14 MW) neutral beam-heated L-mode plasma with $B_T$ = 1.7 T and $I_P$ = 800 kA. The beam power is delivered by a combination of neutral beams with different injection geometries (see e.g. [32]) injecting at different times. For this discharge, the DBS system operated in O-mode polarization and was programmed to step using 1.2 GHz increments starting from 60 GHz at t ~ 370 ms, with a dwell time of ~200 ms at each frequency. The high frequency modes are detected as early as t = 2000 ms, when the operating frequency is 69.6 GHz and continue to be detected nearly continuously until t ~ 4000 ms when the operating frequency is 80.4 GHz. The line average plasma density is ~ 5.2×10$^{13}$ cm$^{-3}$ throughout the period where these modes are observed. GAEs are global modes with long radial wavelengths (typically comparable to the minor radius) and their associated density perturbation modulates the millimeter wave index of refraction [33]. This in turn modulates the optical path length of the millimeter wave radiation from the launcher on its path into plasma and then back to the launcher after scattering from turbulence. Figure 7a shows the spectrum of the phase fluctuations ($\tilde{\phi}$) from the low frequency signal, which are proportional to path length fluctuations, for a later portion of the discharge, with slightly higher line average density ~ 5.3×10$^{13}$ cm$^{-3}$, from t = 3400 ms – 4000 ms. The optical path length of the radiation is given by $\phi/k_0$, where $k_0$ is the millimeter wave vacuum wave number, and $\phi$ is related to in-phase and quadrature signals by $I = A\cos(\phi)$ and $Q = A\sin(\phi)$.

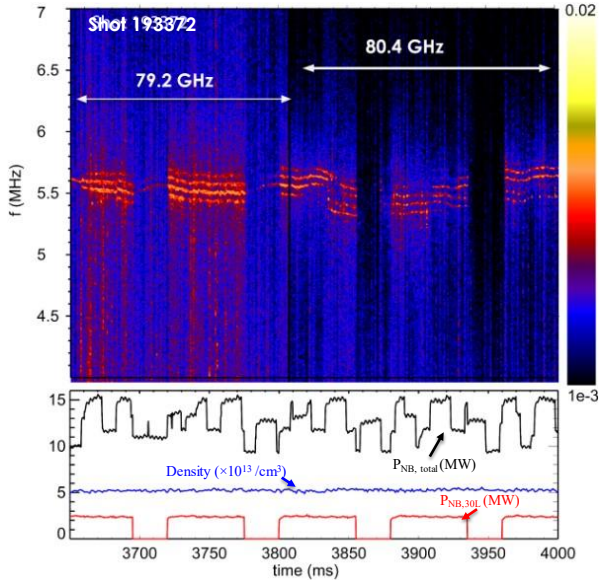

FIG. 7. (a) Phase fluctuation spectrum showing several modes near ~5.5 MHz modes during frequency stepping for an L-mode plasma. (b) Total injected beam power (black line), power injected by "30L" neutral beam (red line), and line averaged $n_e$ (blue line). Note the correlation between the 30L beam injection and the excitation of the ~5.5 MHz coherent modes.

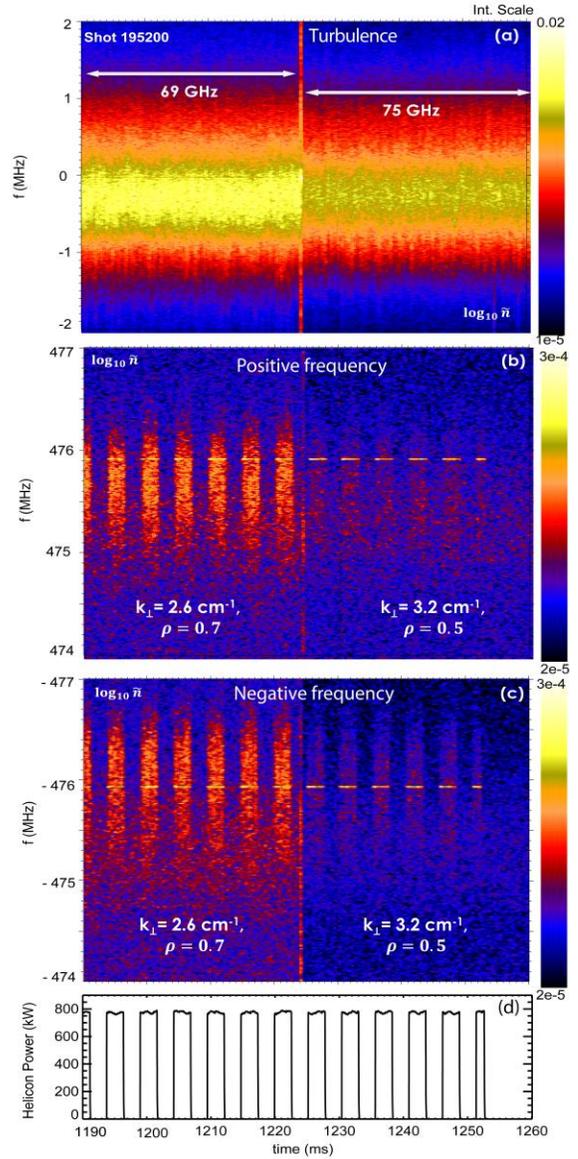

FIG. 8. Illustrating DBS response to helicon wave injection. Quadrature spectra of (a) low frequency turbulence ($\log_{10} \tilde{n}$) signal, (b) positive ($\log_{10} \tilde{n}$) and (c) negative frequency side ($\log_{10} \tilde{n}$) of high frequency broadband helicon signal. (d) helicon injection power measured at klystron output. Two DBS frequency steps 69 GHz to 75 GHz with fixed steering angles and X-mode polarization. Ray tracing indicates DBS shifts from probing fluctuations with $k_\perp$ = 2.6 cm$^{-1}$ at ρ = 0.7 to $k_\perp$ = 3.2 cm$^{-1}$ at ρ = 0.5.

Because of the poloidal aiming and a peak plasma frequency comparable to the operating frequency for this period, DBS probe propagates above the magnetic axis. At 79.2 GHz, it makes the closest approach at $\rho \approx 0.4$, where it probes density fluctuation with $k_\perp \approx 9$ cm$^{-1}$. Figure 7b shows the line average density, and the total neutral beam power, along with the contribution from a particular neutral beam (denoted "30L", located 30° toroidally). Notably, the observed mode activity is destabilized exclusively when the "30L" is injecting, consistent with the wave-particle resonance with beam ion that drives GAEs [32]. Resonance between a beam ion and a wave depends on the wave frequency and wavenumber, as well as the beam ion velocity. Each beam sources a distribution with a narrow range of pitches, $v_{b\parallel}/v_b$, where $v_b$ is beam ion velocity and $v_{b\parallel}$ is the velocity component parallel to the magnetic field. The differing injection geometries of the various beams affect the pitch of the ions they source and thus the waves that the ions can potentially resonantly drive. As discussed in [32], the 30L beam sources a distribution likely to drive waves such as observed here. The frequencies of the modes vary over time within each period of 30L injection and from one period to the next. This is unsurprising, given the evolving plasma conditions, which affect the GAE dispersion relation. However, consistent with the global nature of GAEs, when the operating frequency steps, plasma conditions change little during the settling time and the same mode frequencies are observed before and after the step. This can be seen, for example, at t ≈ 3810 ms in Fig. 7, when the operating frequency steps from 79.2 GHz to 80.4 GHz during a period of 30L injection.

An example of a discharge with helicon injection is shown in Fig. 8. This discharge is characterized by a plasma current ($I_P$) of 1 MA and a toroidal magnetic field ($B_t$) = 1.8 T; the focus of the frequency stepping tests is directed towards simultaneous helicon wave and turbulence measurements, as depicted in Fig. 8. This is a neutral beam heated L-mode plasma. Measurements of density fluctuations in the helicon range of frequency are obtained during a period of constant plasma density with pulsed helicon injection spanning 1190 to 1260 milliseconds. Approximately 800 kW helicon power (measured at klystron output), modulated ON and OFF with a 50% duty cycle at a rate of ~ 194 Hz, is fed into the helicon antenna (Fig. 8d). Two different Doppler Backscattering (DBS) operating frequencies ($f_{DBS}$) are launched successively (both in X-mode) during the discharge. The received signal in the helicon range of frequencies is measured using the down-conversion receiver circuit, illustrated in Fig. 2. The high-frequency quadrature spectrum (Fig. 8b & 8c) shows modulation correlated with the modulation of the injected helicon power, with a sharp peak at the helicon frequency and broadband fluctuations appearing when the helicon power is ON and only background noise being present when the helicon power is OFF. The sharp peak, which is attributable to pickup of stray RF radiation outside the plasma at the helicon source frequency by the measurement circuit (Fig. 2), is discussed below. Notably, the intensity of the broadband fluctuations decreases abruptly around t ≈ 1225 ms, mirroring a sudden drop in the intensity of the low-frequency turbulence (as depicted in Fig. 8a) simultaneously observed in the quadrature spectrum of the low frequency signal from the DBS. This drop is attributed to a change in the probe frequency from 69 to 75 GHz within the DBS system occurring at that time. Raytracing indicates that with the shift of the operating frequency at t ≈ 1225 ms DBS shifts from measuring density fluctuations with $k_\perp$ = 2.6 cm$^{-1}$ in the outer plasma core at $\rho$ = 0.7 to $k_\perp$ = 3.2 cm$^{-1}$ at the mid-radius of the plasma at $\rho$ = 0.5.

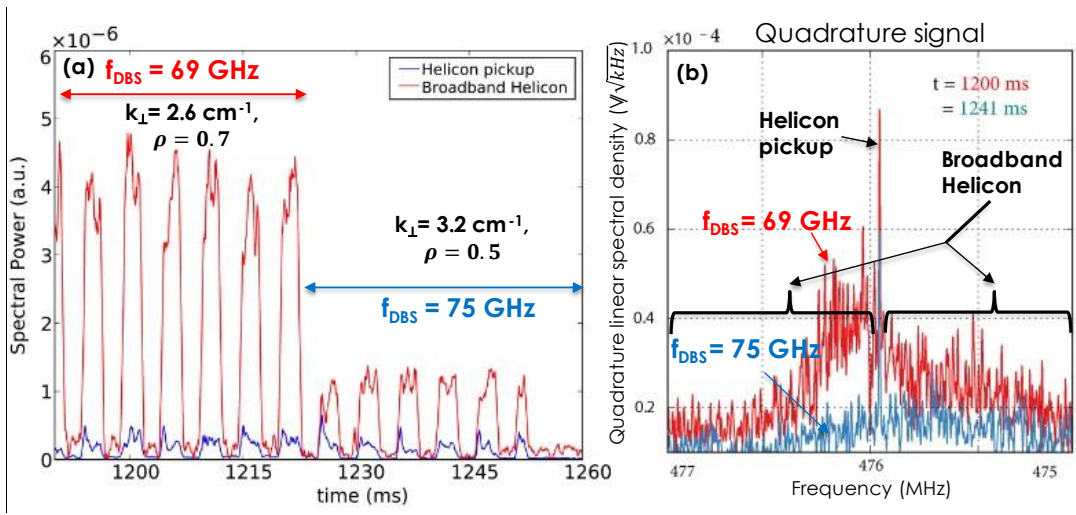

FIG. 9. (a) Spectral power (with background noise subtracted) of helicon pickup peak (~ 476 MHz) and broadband fluctuations around helicon frequency excluding helicon pickup peak during period when operating frequency steps from 69 GHz to 75 GHz. Power in pickup peak shows no response during step, whereas significant change in the broadband fluctuations is observed. (b) Quadrature spectrum (without background noise subtracted) vs frequency at two different times (t = 1200 and 1241 ms) before and after change in operating frequency, illustrating change in broadband power spectra with roughly constant helicon pickup near 476.1 MHz.

Comparison of the broadband helicon fluctuations in Fig. 8b with similar fluctuations in Fig. 8c indicate that the spectrum of broadband fluctuations is not symmetric, leading to an explanation for the cause of the broadband fluctuations. A consistent frequency shift towards more negative values is observed in the broadband fluctuations in the positive and negative helicon frequency range that mirrors a similar shift in the turbulence spectrum. This pattern indicated that the high frequency fluctuations are features of the spectrum of radiation scattered from turbulence rather than radiation scattered from helicon waves. They are sidebands caused by imaging of the spectral feature in the turbulence range of frequencies to frequencies around ± 476 MHz. Two mechanisms are proposed in ref [16] for producing the sidebands. One is path length modulation by density or magnetic field strength perturbations associated with the helicon wave that affects both the probe beam and the probe beam radiation returning along the path to the DBS after backscattering from turbulence. The density and magnetic field strength perturbations perturb the index of refraction of the millimeter waves. This mechanism is not expected to play a role in this case since path length modulation accumulates over the path of the millimeter wave radiation and cancellation occurs when the path crosses through full wavelengths of the perturbation to the index of refraction. The helicon waves have a relatively short wavelength compared to the extent of the region over which path length modulation is accumulated along the millimeter wave ray. From GENRAY ray tracing for the injected helicon wave, the helicon wavelength is predicted to be ~ 4.7 cm when it passes through the DBS measurement region for this case. The extent of the region with significant helicon power along the millimeter wave ray is not well known, but the helicon antenna has a poloidal extent of 20 cm [22], which sets a minimum.

It is expected in this case that the sidebands are caused by the second mechanism proposed in [16], a rapid modulation of the turbulence Doppler shift by the helicon wave E-field in the region where turbulence is measured. Electrons are magnetized for waves at the helicon frequency, and the extremely low frequency turbulence cannot dynamically respond on the time scale of the helicon wave period, so the electron density fluctuations which scatter the millimeter wave probe beam are essentially frozen into the electron fluid and advected by an oscillating $v_{E \times B}$ resulting from the helicon E-field. (Ions are essentially unmagnetized, so they are not advected directly by $E \times B$ drift, but quasi-neutrality ensures ion density fluctuations are driven to match the advected electron density fluctuations.) The spectral analysis of Fig. 8 uses long time records (~ 1/3 msec) so the rapid oscillation of the Doppler shift of the scattered spectrum manifests numerically as high frequency sidebands which are images of spectrum in the low frequency range of turbulence.

A careful analysis of fluctuations in the helicon frequency range (Fig. 9) shows that the broadband fluctuations observed during helicon injection undergo an approximately 75% reduction in mean power when the probe frequency increases from 69 to 75 GHz. Figure 9a shows the mean power fluctuations within the range 475 – 477 MHz, excluding a sharp peak as seen in Fig. 9b at ~ 476 MHz and also excluding background noise. (Note that background noise is not subtracted from the spectra in Fig. 9b.) This decrease occurs as the measurement location shifts away from the outer plasma to the plasma mid-radius even though the peak helicon power to the antenna during each pulse remains nearly constant throughout the period shown. The reduction in the high frequency broadband is consistent with interpretation of the broadband features as sidebands of the low frequency turbulence spectrum since the simultaneously observed turbulence also undergoes a substantial reduction at the same time. However, helicon amplitude at the location of measurement may also change as the measurement location changes, which would play a role. The amplitude of the Doppler shift modulation is expected to affect the ratio of the power in the sideband to that in the turbulence. An analysis of these measurements to infer the helicon amplitude requires further development of a synthetic diagnostic model and is left to future work.

The sharp peak is excluded from the above analysis because it is due to pickup by the measurement circuit (Fig. 2) of stray RF radiation outside the plasma at the helicon source frequency. The contribution of pickup is confirmed by blocking the beam from entering the plasma for a similar discharge. The peak is still observed in the measured spectrum, even as the spectral power is reduced to noise levels throughout the rest of the spectrum. Note that the measured frequency of the peak, which can be seen in Fig. 9b, is shown to be ~ 476.1 MHz, slightly different from true value of 476 MHz produced by the helicon source. This is an artifact of how the measured frequencies are adjusted to account for the downshifting in the measurement circuit using the 500 MHz crystal oscillator (Fig. 2). There is an uncorrected small temperature-related deviation of the crystal oscillator frequency from the rated value of 500 MHz. The mean power of the pickup peak (with background noise subtracted) is shown separately in Fig. 9a. Notably, the mean power of peak remains unchanged, independent of the external DBS launch frequency ($f_{DBS}$). In principle, a peak at the helicon source frequency could come from the millimeter wave radiation directly backscattered from the helicon wave injected into the plasma. However, the insensitivity of the power in the peak to measurement location, in this case, indicates that it is entirely attributable to RF pickup since a change in measurement location should also result in a change in any millimeter wave power scattered directly from the injected helicon wave.

The ECH launcher steering capability is illustrated in Fig. 10. A high-power, neutral beam-heated H-mode plasma ($I_P$ = 1.1 MA and $B_T$ = 2.02 T) is selected as the target discharge for the test. Using a fixed frequency of 73.8 GHz with X-mode polarization, a wavenumber scan is performed at a measurement location of $\rho \approx 0.97$. The steerable mirror is programmed before the discharge to scan through a range of poloidal angles at a constant 0° toroidal angle to investigate low-k fluctuations, as illustrated in Fig. 10. The time dependent quadrature spectrum of the low frequency signal in the top panel (Fig, 10a) shows turbulent density fluctuations, while the requested and actual mirror steering angles (measured relative to normal incidence) are shown in the bottom panel, along with the line average density. Significant variations can be observed in intensity and width of the spectrum as the mirror position is adjusted, particularly when the mirror is steered to probe very low k. Figure 10b shows the requested poloidal and actual angle poloidal steering angle of the mirror. The value shown is relative to the angle needed for normal incidence, which is determined from ray tracing. Density is constant throughout the period shown in the figure, as can be seen in Fig 10b. Two full scans spanning a nearly 8° range of poloidal angles ($k_\perp \sim 3 - 7$ cm$^{-1}$) are shown in the figure during the period t $\approx$ 1800 – 2550 ms and the spectrum of fluctuations can be seen to exhibit a similar dependence on poloidal angle in each scan, suggesting that the turbulent spectrum is approximately stationary during this period. A wavenumber power spectrum for three different probed wavenumbers is plotted in Fig. 10c showing an often-observed trend of decreasing fluctuation power with increasing k [34]. This power spectrum has been calculated from the quadrature spectrum integrated over +/- 1MHz frequency band. As can be seen, there is a delay of approximately 40 ms between the request and the achieved mirror angle in the motor feedback system. This delay is attributed to a combination of communication latency and the inertia involved in initiating mirror movement from a static position.

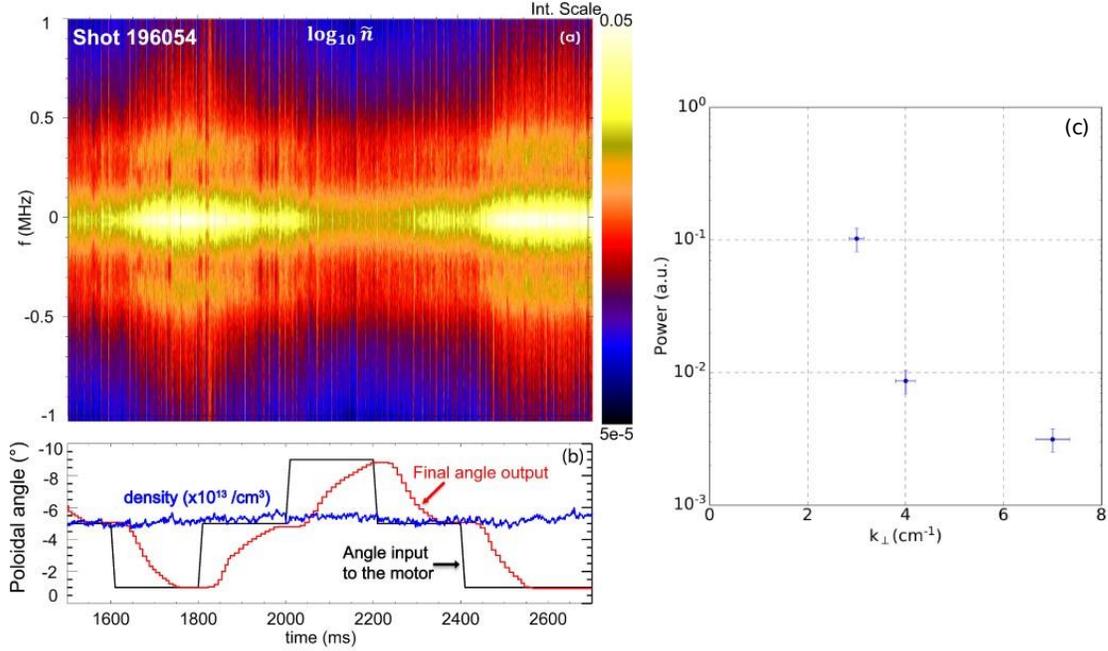

FIG.10. (a) Illustrating in-shot scan of poloidal steering angle and corresponding changes in quadrature turbulence spectrum. (b) Showing constant density, requested and resulting beam poloidal steering angles. 0º corresponds to normal incidence and negative angles are downward relative to 0º. (c) Total quadrature power within frequency range $|f| \leq 1$ MHz for three different poloidal steering angles shown vs. probed wavenumber.

## 4. CONCLUSIONS

New frequency stepping and real-time steering capabilities have been demonstrated in DIII-D plasmas for a DBS system that measures turbulence and RF waves simultaneously. The system has been integrated into a steerable ECH launcher and uses a programmable synthesizer-based millimeter-wave source. This integration addresses technical challenges and sets a model for future burning plasma facilities with ECH. The steering capability allows for probing plasma fluctuations across a wide range of wavenumbers within a single discharge. Frequency stepping tests revealed challenges in spatial distribution measurements, suggesting future work on synchronizing steering and frequency stepping to maintain constant wavenumber measurements. The system will also enhance DBS measurement capabilities for DIII-D, extending the capability for turbulence measurements with wave-number matching (i.e. $k_\parallel \sim 0$ at cutoff) to high k ($k_\perp >$ 15cm$^{-1}$) [35] allowing for poloidal correlation studies with other millimeter wave scattering systems that are poloidally separated but at the same toroidal location [36]. Additionally, results presented here reveal broadband helicon fluctuations caused by a mechanism different from that previously reported in Ref. [16]. High-power helicon wave measurements showed high-frequency broadband features in the density fluctuation spectrum near the helicon frequency, believed to be sidebands of turbulence modulated by the helicon wave.

## ACKNOWLEDGMENTS

This material is based upon work supported by the U.S. Department of Energy, Office of Science, Office of Fusion Energy Sciences, using the DIII-D National Fusion Facility, a DOE Office of Science user facility, under Awards DE-FC02-04ER54698, DE-SC0020649 and DE-SC0020337. The authors would like to thank Larry Bradley and the DIII-D Diagnostic team for their engineering as well as technical support during the installation of the DBS diagnostic system.

## AUTHOR DECLARATIONS

### Conflict of Interest

The authors have no conflicts of interest to disclose.

## DATA AVAILABILITY

The data that support the findings of this study are available from the corresponding author upon reasonable request.

## REFERENCES

[1] A. P. Smirnov and R.W. Harvey, Bull. Am. Phys. Soc. 40, 1837 (1995). https://www.compxco.com/genray.html

[2] C. Lau, E.F. Jaeger, N. Bertelli, L.A. Berry, D.L. Green, M. Murakami, J.M. Park, R.I. Pinsker and R. Prater, Nuclear Fusion 58 066004 (2018). https://doi.org/10.1088/1741-4326/aab96d

[3] E. Mazzucato, Rev Sci Instrum 69, 2201–2217 (1998). https://doi.org/10.1063/1.1149121

[4] C. Laviron, A. J. H. Donné, M. E. Manso and J. Sanchez, Plasma Phys. Control. Fusion 38, 905–936, (1996). https://doi.org/10.1088/0741-3335/38/7/002

[5] G. D. Conway, Plasma Phys. Control. Fusion 46, 951–970 (2004). https://doi.org/10.1088/0741-3335/46/6/003

[6] G. D. Conway, Nucl. Fusion 46, S799 (2006). https://doi.org/10.1088/0029-5515/46/9/S01

[7] G. D. Conway, Plasma Phys. Control. Fusion 50, 124026 (2008). https://doi.org /10.1088/0741-3335/50/12/124026

[8] T. L. Rhodes, W. A. Peebles, X. Nguyen, M. A. VanZeeland, J. S. Degrassie, E. J. Doyle, G. Wang and L. Zeng, Rev. Sci. Instrum. 77, 10E922-1-8 (2006). https://doi.org/10.1063/1.2235874

[9] W. A. Peebles, T. L. Rhodes, J. C. Hillesheim, L. Zeng and C. Wannberg, Rev. Sci. Instrum. 81, 10 10D902 (2010). https://doi.org/10.1063/1.3464266

[10] J. C. Hillesheim, W. A. Peebles, T. L. Rhodes, L. Schmitz, T. A. Carter, P. A. Gourdain, and G. Wang, Rev Sci Instrum 80, 083507 (2009). https://doi.org/10.1063/1.3205449

[11] R. G. L. Vann, K. J. Brunner, R. Ellis, G. Taylor, and D. A. Thomas, Rev Sci Instrum 87, 11D902 (2016). https://doi.org/10.1063/1.4962253

[12] N A Crocker, W A Peebles, S Kubota, J Zhang, R E Bell, E D Fredrickson, N N Gorelenkov, B P LeBlanc, J E Menard, M Podest`a, S A Sabbagh, K Tritz and H Yuh, Plasma Phys. Control. Fusion 53, 105001 (2011). http://dx.doi.org/10.1088/0741-3335/53/10/105001

[13] J.C. Hillesheim, N.A. Crocker, W.A. Peebles, H. Meyer, A. Meakins, A.R. Field, D. Dunai, M. Carr, N. Hawkes and the MAST Team, Nucl. Fusion 55, 073024, (2015). http://dx.doi.org/10.1088/0029-5515/55/7/073024

[14] T. L. Rhodes, C. A. Michael, P. Shi, R. Scannell, S. Storment, Q. Pratt, R. Lantsov, I. Fitzgerald, V. H. Hall-Chen, N. A. Crocker, and W. A. Peebles, Rev. Sci. Instrum. 93, 113549 (2022). https://doi.org/10.1063/5.0101848

[15] N.A. Crocker, S.X. Tang, K.E. Thome, J.B. Lestz, E.V. Belova, A. Zalzali, R.O. Dendy, W.A. Peebles, K.K. Barada, R. Hong et al., Nuclear Fusion 62:2, 026023 (2022). https://doi.org/10.1088/1741-4326/ac3d6a

[16] S. Chowdhury, N. A. Crocker, W. A. Peebles, T. L. Rhodes, L. Zeng, R. Lantsov, B. Van Compernolle, M. Brookman, R. I. Pinsker, C. Lau, Rev. Sci. Instrum. 94, 073504 (2023); https://doi.org/10.1063/5.0149654

[17] S. Shiraiwa, S. Baek, A. Dominguez, E. Marmar, R. Parker and G. J. Kramer, Rev. Sci. Instrum. 81, 10D936 (2010). https://doi.org/10.1063/1.3492370

[18] C. Lau, M. Brookman, A. Dimits, B. Dudson, Elijah Martin, Robert I. Pinsker, Matt Thomas and Bart Van Compernolle, Nucl. Fusion 61 126072 (2021). https://doi.org/10.1088/1741-4326/ac36f3

[19] M. Cengher, J. Lohr, Y. Gorelov, A. C. Torrezan, D. Ponce, X. Chen and C. Moeller, IEEE Trans Plasma Sci, 44, pp. 3465-3470 (2016). https://doi.org/10.1109/tps.2016.2542789

[20] M. Cengher, X. Chen, R. Ellis, Y. Gorelov, J. Lohr, C. Moeller, D. Poncea and A. C. Torrezan, Fusion Eng. Des. 123, 295–298 (2017). https://doi.org/10.1016/j.fusengdes.2017.05.022

[21] R.I. Pinsker, R. Prater, C.P. Moeller, J.S. deGrassie, C.C. Petty, M. Porkolab, J.P. Anderson, A.M. Garofalo, C. Lau, A. Nagy et al., Nucl. Fusion 58 106007 (2018). https://doi.org/10.1088/1741-4326/aad1f8

[22] B. Van Compernolle, M.W. Brookman, C.P. Moeller, R.I. Pinsker, A.M. Garofalo, R. O'Neill, D. Geng, A. Nagy, J.P. Squire, K. Schultz et al., Nucl. Fusion 61 116034 (2021). https://doi.org/10.1088/1741-4326/ac25c0

[23] R.I. Pinsker, B. Van Compernolle, S.X. Tang, J.B. Lestz, C.P. Moeller, C.C. Petty, A. Dupuy, J.P. Squire, A.M. Garofalo, M. Porkolab, J.C. Rost, S.G. Baek, A. Nagy, S. Chowdhury, N.A. Crocker, G.H. Degrandchamp, A.G. Mclean, K.R. Gage, A. Marinoni, E.H. Martin, G. Ronchi and the

DIII-D Team, 'First high-power helicon results from DIII-D' ID: IAEA-CN-316-2111 (Submitted to Nucl. Fusion, 2024), 29th IAEA Fusion Energy Conference (FEC2023).

[24] R. Callis, W. Cary, C. Moeller, R. Freeman, R. Prater, D. Remsen, and L. Sevier, AIP Conf. Proc. 244, 24–27 (1992). https://doi.org/10.1063/1.41658

[25] R. Prater, D. Farina, Yu. Gribov, R.W. Harvey, A.K. Ram, Y.-R. Lin-Liu, E. Poli, A.P. Smirnov, F. Volpe, E. Westerhof, A. Zvonkov and the ITPA Steady State Operation Topical Group, Nucl. Fusion 48, 035006 (2008). http://dx.doi.org/10.1088/0029-5515/48/3/035006

[26] General atomics waveguide switch: https://www.ga.com/microwave-technologies/

[27] Anderson, J., Doane, J., Moeller, C., Grunloh, H., O'Neill, R., Brookman, M., Smiley, M. and Su, D., 2019. Design and Performance of Microwave Components for ECH and ECE Applications at General Atomics. In EPJ Web of Conferences (Vol. 203, p. 04001). EDP Sciences. https://doi.org/10.1051/epjconf/201920304001

[28] Goldsmith, Paul F. "Quasi-optical techniques." Proceedings of the IEEE 80, no. 11 (1992): 1729-1747. https://doi.org/10.1109/5.175252

[29] Kowalski, E. J., et al. IEEE Trans. Plasma Sci. 42, 29–37 (2014). https://doi.org/10.1109/TPS.2013.2288493

[30] B. W. Rice, K. H. Burrell, L. L. Lao, and Y. R. Lin-Liu, Phys. Rev. Lett. 79, 2694 (1997). https://doi.org/10.1103/PhysRevLett.79.2694

[31] C. Chrystal, S. R. Haskey, K. H. Burrell, B. A. Grierson, and C. S. Collins, Rev. Sci. Instrum. 92, 043518 (2021); https://doi.org/10.1063/5.0043087

[32] S. X. Tang, T. A. Carter, N. A. Crocker, W. W. Heidbrink, J. B. Lestz, R. I. Pinsker, K. E. Thome, M. A. Van Zeeland and E. V. Belova, Phys. Rev. Lett. 126, 155001 (2021). https://doi.org/10.1103/physrevlett.126.155001

[33] L. Zeng, N. A. Crocker, T. L. Rhodes, and W. A. Peebles, Rev. Sci. Instrum. 92, 043550 (2021); https://doi.org/10.1063/5.0043121

[34] Q. Pratt, V. Hall-Chen, T.F. Neiser, R. Hong, J. Damba, T.L. Rhodes, K.E. Thome, J. Yang, S.R. Haskey, T. Cote and T. Carter, Nucl. Fusion 64 016001 (2024); https://doi.org/10.1088/1741-4326/ad0906

[35] J. Damba, Q. Pratt, V. H. Hall-Chen, R. Hong, R. Lantsov, R. Ellis, T. L. Rhodes, Rev. Sci. Instrum. 93, 103549 (2022); https://doi.org/10.1063/5.0101864

[36] T. L. Rhodes, R. Lantsov, G. Wang; R. Ellis, W. A. Peebles, Rev. Sci. Instrum. 89, 10H107 (2018); https://doi.org/10.1063/1.5035427